\documentclass[11pt]{article}
\usepackage{graphicx}
\usepackage[breaklinks]{hyperref}
\usepackage{booktabs}
\usepackage{amsmath,amssymb}

\begin{document}
\title{Phenomenology of Quantum Gravity and its Possible Role in Neutrino Anomalies}

\author{Mario A. Acero$^{1,2}$\footnote{marioacero@mail.uniatlantico.edu.co} \ and Yuri Bonder$^{2,3}$\footnote{ybonder@indiana.edu}}
\date{\begin{small}
       $^1$ Departamento de F\'isica, Universidad del Atl\'antico,\\Km 7 antigua v\'ia a Puerto Colombia, Barranquilla, Colombia.\\ \vspace{0.3cm}
$^2$ Instituto de Ciencias Nucleares,\\Universidad Nacional Aut\'onoma de M\'exico,\\ Apartado Postal 70-543, M\'exico D.~F., 
04510, M\'exico.\\\vspace{0.3cm}
$^3$ Physics Department, Indiana University,\\ Bloomington, Indiana, 47405, U.S.A.\\
(Dated: IUHET 570, September 2012; accepted in Edition Open Access)
      \end{small}}

\maketitle

\begin{abstract}
New phenomenological models of Quantum Gravity have suggested that a Lorentz-Invariant discrete spacetime structure may become manifest through a nonstandard coupling of matter fields and spacetime curvature. On the other hand, there is strong experimental evidence suggesting that neutrino oscillations cannot be described by simply considering neutrinos as massive particles. In this manuscript we motivate and construct one particular phenomenological model of Quantum Gravity that could account for the so-called neutrino anomalies.
\end{abstract}

\section{Introduction}

To construct a theory that reconciles Quantum Mechanics and General Relativity is one of the most challenging problems in Physics. 
This still unfinished theory is called Quantum Gravity (QG), and we believe that the difficulty in building such theory may be, in part, by the lack 
of experimental guidance. Regarding particle physics, the Standard Model of particles (SM) includes three massless neutrinos. The Higgs mechanism \cite{Higgs:1964ia}, through which the masses of all other fermions (as well as bosons) are generated, does not apply for neutrinos because the neutrino fields do not have right handed components \cite{Giunti}. However, the experimental observation of neutrinos changing from a flavor to another, a phenomenon known as \emph{neutrino oscillation}, has motivated people to suggest that neutrinos are actually massive and current research looks for an extension of the SM to include neutrino masses. The simplest extension of the SM (but certainly not the only one, see Refs.~\cite{Giunti,Zuber,Mohapatra:1998rq}) is to include right handed components of the neutrino fields, so that they acquire mass through the same mechanism as the other particles.

Nevertheless, to include neutrino masses seems to be insufficient to account for all observations. In this work we argue that the anomalous neutrino oscillations could be regarded as traces of the quantum nature of gravity. More concretely, we propose a modification to the simple extension of the SM described above motivated by a phenomenological model of QG in order to explain all neutrino observations. Before we continue, we warn the reader that we only present the motivation and possible applications of a class of phenomenological models of QG to neutrinos; a deeper study of this issue is needed to test if these models are a feasible explanation of the neutrino anomalies.

\section{Neutrino oscillations}

Neutrino oscillations are transition of a neutrino in a definite flavor state
into a neutrino with a different flavor. The basic idea is that a neutrino flavor state is a linear combination of states with definite mass. The \emph{oscillation probability}, in the two--neutrinos approximation, is given by \cite{Giunti}
\begin{equation}\label{eq:OscProb}
 P_{\nu_{\alpha} \to \nu_{\beta}}(L,E) = \sin^2(2\theta) \sin^2\left(\frac{\Delta m^2 L}{4E}\right), \quad \alpha \neq \beta,
\end{equation}
where $L$ and $E$ are, respectively, the distance traveled by the neutrino and its energy (both in the laboratory reference frame), and the two fundamental parameters of this process are the \emph{mixing angle} $\theta$ and the \emph{masses--squared difference} $\Delta m^2 \equiv m_2^2 - m_1^2$. It is under this effective model that most of the experimental data have been analyzed, given that many experiments are not sensitive to the effects of three-neutrino mixing \cite{Giunti}. 

Now, from observations of solar, atmospheric, reactor, and accelerator--based neutrino--oscillation experiments, it has been possible to establish firmly the existence of three mixing angles and two separated mass--splitting parameters of order $10^{-5}\rm{eV}^2$ and $10^{-3} \rm{eV}^2$ (for an updated combined data analysis, see \cite{Fogli:2012ua,GonzalezGarcia:2010er,Schwetz:2008er}). Yet, there are some results that cannot be explained with these parameters. The Liquid Scintillator Neutrino Detector (LSND) \cite{Athanassopoulos:1997pv,Aguilar:2001ty} experiment found that their oscillation data point to $\Delta m^2=O(1\,\rm{eV}^2)$, which is much larger than those $\Delta m^2$ found by the experiments mentioned before \cite{Eguchi:2002dm,Greenwood:1996pb,Michael:2006rx,Wendell:2010md}. More recently, in an attempt to check this anomalous outcome, the MiniBooNE collaboration \cite{AguilarArevalo:2009xn,AguilarArevalo:2010wv} found that, with a $99\%$ confidence level, their analysis leads to a $\Delta 
m^2$ that is consistent with that from LSND. Note that both experiments, LSND and MiniBooNE, produced the neutrinos in accelerators and have the same $L/E$ (see equation (\ref{eq:OscProb})). Additional anomalous results have been under study and include the Reactor antineutrino anomaly \cite{Mention:2011rk} and the Gallium anomaly \cite{Giunti:2006bj,Acero:2007su}.

Currently, a great effort is done to clarify these issues, both from the theoretical and experimental point of view and a number of experiments currently running, and different proposals for the future, are devoted to it \cite{KlapdorKleingrothaus:1994kq,Lobashev:2003kt,Drexlin:2005zt,Ghoshal:2010wt,Aseev:2011dq,Abazajian:2011dt,Blennow:2012gj}. On theoretical grounds, perhaps the most popular explanation is the existence of, at least, one additional neutrino which has to have different properties compared to those included in the SM. This (or these) new neutrino is known as \emph{sterile}, given that it does not take part in the weak interactions of the SM \cite{Giunti}. However, there is no further evidence supporting the existence of sterile neutrinos. In this work, we take a different strategy where there is no need to add new particles. In contrast, we propose that gravity, whose fundamental version is still unknown, may couple to the neutrino fields in a non-standard way, producing the anomalous 
neutrino oscillations. In addition, if gravity is behind neutrino oscillations, it is conceivable that these depend on the gravitational environment, as is suggested by the aforementioned experimental results. In the next section, we briefly present the phenomenological model of QG that gives rise to these couplings.

\section{Lorentz Invariant Phenomenology of Quantum Gravity}

The phenomenology of QG has been dominated, in the last years, by searching for Lorentz-Invariance (LI) violations. This may be motivated by the fact that a naive discrete spacetime structure naturally selects preferred directions. Besides the 
significant empirical bounds on LI violations (for the most complete collection of bounds see \cite{Bounds}), Collins 
\textit{et al.} \cite{Collins} have argued that a LI violating discrete spacetime inhabited by quantum fields can be discarded by experiments. Essentially, the radiative corrections would magnify the effects of a LI violating discrete spacetime up to the point where they should have been observed. These arguments motivated a new type of phenomenological models of QG \cite{QGP1,QGP2} where a LI discrete spacetime structure is sought precisely by using the hypothesis that the 
symmetry of spacetime building blocks is LI.

It is hard to envision a discrete spacetime structure that respects LI. However, in order to build a phenomenological model, there is no need to have a concrete picture of such a structure. The basic idea is that the presence of a LI discrete spacetime structure may reveal itself when there is a mismatch between the symmetries of spacetime (at a macroscopic scale) and those symmetries of its building blocks. As mentioned above, in these models one assumes that spacetime building blocks are LI, thus, the mismatch with the macroscopic symmetry would occur when the macroscopic spacetime is not LI. This, in turn, happens in curved spacetime regions, suggesting that the effects of a LI discrete spacetime structure could manifest themselves as non-standard couplings of curvature and matter fields.

Studying a coupling of matter and the Ricci tensor (or the curvature scalar) is not interesting phenomenologically because, 
according to Einstein's equations, these geometrical objects at a given spacetime point are determined by the matter at that 
same point. Thus, to couple matter with the Ricci tensor can be considered at the phenomenological level as a self-coupling. 
Thus, the Weyl tensor $W_{abcd}$, which loosely speaking is the part of the Riemann tensor that remains when the Ricci part is 
subtracted \cite{Wald}, is the object that should be coupled with the matter fields. Moreover, the coupling must vanish in flat 
spacetime regions where spacetime is actually LI.

In the past, one particular model was extensively studied. It involves fermionic matter fields that couple to the eigenvalues 
and eigenvectors of two Hermitian operators built out of the Weyl tensor through complicated couplings \cite{QGP2}. This model has been able to produce some bounds in the neutron sector \cite{QGP3} and to motivate an experiment where the effect predicted by the model was sought \cite{QGP4,Adelberger} and bounds on the electron sector were obtained. In the next section, a particular model for neutrinos that may help explain some of the anomalies described above is presented.

\section{Neutrino effective mass}

For simplicity, we only consider Dirac neutrinos with non-vanishing masses and right handed components. The strategy is to generate effective masses that depend on the gravitational environment. Following the motivation discussed above, this effective mass should be obtained 
through non-minimal couplings of spacetime curvature (Weyl tensor) and the neutrino fields. This coupling should vanish in flat 
spacetime regions and must respect gauge invariance to have a theory with a well posed initial value formulation (see the discussion on that matter in Ref.~\cite{Wald}.)

To define this coupling term we write the Lagrangian density describing massive Dirac massive neutrinos in a curved background:
\begin{eqnarray}
\mathcal{L}_{g+ew}&=& ie \bar{\nu}_{L\alpha} e_\mu^a \gamma^\mu D^{(g+ew)}_a \nu_{L\alpha} + ie
\bar{\nu}_{R\alpha} e_\mu^a \gamma^\mu D^{(g)}_a \nu_{R\alpha} \nonumber\\
&&\qquad - e \Gamma_{\alpha \beta} \left(\bar{\nu}_{L\alpha}\phi \nu_{R\beta} +
\bar{\nu}_{R\alpha} \phi \nu_{L\beta}\right),\label{L1}
\end{eqnarray}
where $\nu_L$ and $\nu_R$ are the left and right neutrinos, $\phi$ is (one component of) the
Higgs field, $\Gamma_{\alpha\beta}$ are the (dimensionless) Yukawa coupling constants, $e_\mu^a$ are the tetrads, $e$ is the spacetime natural volume form and $D^{(g+ew)}_a$ is the covariant derivative 
including gauge interaction and gravity while $D^{(g)}_a$ contains only the gravitational part. The indices $\alpha,\beta$ label the 
neutrino flavor. The charged lepton part of the Lagrangian, which should be included to have explicit gauge invariance, is not written since the gauge interaction is not considered in what follows.

To respect gauge invariance,
the gravity modification to the mass term must enter into the Lagrangian
density (\ref{L1}) though the replacement
\begin{equation}
\Gamma_{\alpha\beta}\rightarrow
\Gamma_{\alpha\beta}+b_{\alpha\beta}f\left(\frac{W}{M_P^2}\right),
\end{equation}
where $b_{\alpha\beta}$ are the coupling coefficients, $W(x) \equiv \sqrt{W_{abcd} W^{abcd}}$ and $f$ is a dimensionless real function. The Planck mass, $M_P$, is introduced 
in such a way that the argument of $f$ is also dimensionless. The simplest function $f$ that is only suppressed by one power of 
$M_P$ in the denominator is 
\begin{equation}
f\left(\frac{W}{M_P^2}\right)=\frac{\sqrt{W}}{M_P},
\end{equation}
which is the function we consider.

Once the gauge symmetry is spontaneously broken, the gravitational part of the Lagrangian density (\ref{L1}) takes the form
\begin{eqnarray}
\mathcal{L}_g&=& ie \bar{\nu}_{L\alpha} e_\mu^a \gamma^\mu D^{(g)}_a \nu_{L\alpha} + ie
\bar{\nu}_{R\alpha} e_\mu^a \gamma^\mu D^{(g)}_a \nu_{R\alpha} \nonumber\\
&&\qquad - e \left(m_{\alpha\beta}+a_{\alpha\beta}\frac{\sqrt{W}}{M_P}\right)
\left(\bar{\nu}_{L\alpha} \nu_{R\beta} + \bar{\nu}_{R\alpha} \nu_{L\beta}\right), \label{L 2}
\end{eqnarray}
where $m_{\alpha\beta}=\Gamma_{\alpha \beta} <\phi> $ and $a_{\alpha\beta}=b_{\alpha \beta} <\phi>$. Observe that the mass matrix in this case is
\begin{equation}
M_{\alpha\beta}(x)\equiv m_{\alpha\beta}+a_{\alpha\beta}\frac{\sqrt{W(x)}}{M_P}.
\end{equation}
Typically, $m_{\alpha\beta}$ generates neutrino flavor mixing. In the case we are dealing with, these oscillations would be caused by $M_{\alpha\beta}$ which, in all cases of phenomenological interest can be thought as $ m_{\alpha\beta}$ plus a small modulation that depends on the gravitational environment. As neutrinos from different sources (solar, atmospheric, reactor, and accelerator) travel in different gravitational backgrounds, according to the model presented here, they should oscillate slightly differently.

In order to gain some intuition, we consider the effects of this model for neutrinos traveling closely to the Earth's surface, as it happens in accelerator and reactor experiments. In this case $W$ can be taken approximately as constant given by $W=\sqrt{48} M/R^3$ where $R$ and $M$ stand for the radius and mass of the Earth, respectively. The numerical value is $\sqrt{W} \approx  10^{-46} M_P$, which would then require $a_{\alpha \beta}$ to be extremely large in order to produce any measurable effect. Thus, at first sight this model seems to be ruled out. However, let us remind the reader that the size of $a_{\alpha \beta}$ is somehow artificial since we put $M_P$ by hand. Moreover, a different function $f$ could be chosen that could make testable predictions. In any case, a much deeper analysis is required. In particular, one would need to try to fit the free parameters of the model to explain the neutrino anomalies before taking these models seriously. This may be particularly difficult to achieve because,
 in certain cases, the tidal effects of a wall or a mountain can dominate over the effects the entire Earth (see Ref. \cite{Landau}), thus, a very precise knowledge of the gravitational--source distribution on the neutrinos path may be needed to correctly model the neutrino oscillations. Also, the effects of matter, which also generate neutrino oscillation \cite{matter}, must be considered. An intriguing possibility is to try to mimic the well--know MSW \cite{MSW1,MSW2} effect and search for gravitational environments where resonances could be expected\footnote{We thank R. Lehnert for this suggestion.}.

To conclude, we want to stress the reasons that motivated us to consider gravity as a possible explanation for the anomalous neutrino behavior. First, we know gravity exists, thus, we do not need to invoke new fields/particles that have not been observed to account for the anomalies. Second, it is conceivable that QG may influence matter in exotic ways and these effects could become manifest at scales below the Planck regime. Third, neutrino experiments are done with particles that have traveled in different gravitational environments, which may account, at least in part, for the different behavior.

\section*{Acknowledgements}
We want to thank D. Sudarsky, J.S. D\'iaz, R. Lehnert, and V.A. Kosteleck\'y for their comments. MAA acknowledges support by the Mexican RedFAE, CONACyT, during the initial development of this work. YB was partially supported by the research grants CONACyT 101712, PAPIIT-UNAM IN107412 and by the Department of Energy under grant number DE-FG02-91ER40661 and also by the Indiana University Center for Spacetime Symmetries. 

\bibliographystyle{unsrt}
\bibliography{Contribution}

\begin{thebibliography}{10}

\bibitem{Higgs:1964ia}
P.W. Higgs.
\newblock {\em Phys.Lett.}, 12:132, 1964.

\bibitem{Giunti}
C.~Giunti and C.W. Kim.
\newblock {\em Fundamentals of Neutrino Physics and Astrophysics}.
\newblock Oxford University Press, 2007.

\bibitem{Zuber}
K.~Zuber.
\newblock {\em Neutrino Physics}.
\newblock Series in High Energy Physics, Cosmology and Gravitation, Taylor \&
  Francis Group, 2004.

\bibitem{Mohapatra:1998rq}
R.N. Mohapatra and P.B. Pal.
\newblock {\em Massive neutrinos in physics and astrophysics}.
\newblock World Scientific Publisher Co. Pte. Ltd., 1998.

\bibitem{Fogli:2012ua}
G.L. Fogli et~al.
\newblock {\em Phys.Rev.}, D86:013012, 2012.

\bibitem{GonzalezGarcia:2010er}
M.C. Gonzalez-Garcia, M.~Maltoni, and J.~Salvado.
\newblock {\em JHEP}, 1004:056, 2010.

\bibitem{Schwetz:2008er}
T.~Schwetz, M.A. Tortola, and J.W.F. Valle.
\newblock {\em New J.Phys.}, 10:113011, 2008.

\bibitem{Athanassopoulos:1997pv}
C.~Athanassopoulos et~al.
\newblock {\em Phys.Rev.Lett.}, 81:1774, 1998.

\bibitem{Aguilar:2001ty}
A.A. Aguilar-Arevalo et~al.
\newblock {\em Phys.Rev.}, D64:112007, 2001.

\bibitem{Eguchi:2002dm}
K.~Eguchi et~al.
\newblock {\em Phys.Rev.Lett.}, 90:021802, 2003.

\bibitem{Greenwood:1996pb}
Z.D. Greenwood et~al.
\newblock {\em Phys.Rev.}, D53:6054, 1996.

\bibitem{Michael:2006rx}
D.G. Michael et~al.
\newblock {\em Phys.Rev.Lett.}, 97:191801, 2006.

\bibitem{Wendell:2010md}
R.~Wendell et~al.
\newblock {\em Phys.Rev.}, D81:092004, 2010.

\bibitem{AguilarArevalo:2009xn}
A.A. Aguilar-Arevalo et~al.
\newblock {\em Phys.Rev.Lett.}, 103:111801, 2009.

\bibitem{AguilarArevalo:2010wv}
A.A. Aguilar-Arevalo et~al.
\newblock {\em Phys.Rev.Lett.}, 105:181801, 2010.

\bibitem{Mention:2011rk}
G.~Mention et~al.
\newblock {\em Phys.Rev.}, D83:073006, 2011.

\bibitem{Giunti:2006bj}
C.~Giunti and M.~Laveder.
\newblock {\em Mod.Phys.Lett.}, A22:2499, 2007.

\bibitem{Acero:2007su}
M.A. Acero, C.~Giunti, and M.~Laveder.
\newblock {\em Phys.Rev.}, D78:073009, 2008.

\bibitem{KlapdorKleingrothaus:1994kq}
H.V. Klapdor-Kleingrothaus.
\newblock {\em Prog.Part.Nucl.Phys.}, 32:261, 1994.

\bibitem{Lobashev:2003kt}
V.M. Lobashev.
\newblock {\em Nucl.Phys.}, A719:153, 2003.

\bibitem{Drexlin:2005zt}
G.~Drexlin.
\newblock {\em Nucl.Phys.Proc.Suppl.}, 145:263--267, 2005.

\bibitem{Ghoshal:2010wt}
P.~Ghoshal and S.T. Petcov.
\newblock {\em JHEP}, 1103:058, 2011.

\bibitem{Aseev:2011dq}
V.N. Aseev et~al.
\newblock {\em Phys.Rev.}, D84:112003, 2011.

\bibitem{Abazajian:2011dt}
K.N. Abazajian et~al.
\newblock {\em Astropart.Phys.}, 35:177, 2011.

\bibitem{Blennow:2012gj}
M.~Blennow and T.~Schwetz.
\newblock {\em JHEP}, 1208:058, 2012.

\bibitem{Bounds}
V.A. Kosteleck\'y and N.~Russell.
\newblock {\em Rev. Mod. Phys.}, 83:11, 2011.

\bibitem{Collins}
J.~Collins et~al.
\newblock {\em Phys. Rev. Lett.}, 93:191301, 2004.

\bibitem{QGP1}
A.~Corichi and D.~Sudarsky.
\newblock {\em Int. J. Mod. Phys.}, D 14:1685, 2005.

\bibitem{QGP2}
Y.~Bonder and Sudarsky D.
\newblock {\em Class. Quantum Grav.}, 25:105017, 2008.

\bibitem{Wald}
R.M. Wald.
\newblock {\em General Relativity}.
\newblock University of Chicago Press, 1984.

\bibitem{QGP3}
Y.~Bonder and Sudarsky D.
\newblock {\em Rep. Math. Phys.}, 64:169, 2009.

\bibitem{QGP4}
Y.~Bonder and Sudarsky D.
\newblock {\em AIP Conf. Proc.}, 1256:157, 2010.

\bibitem{Adelberger}
W.A. Terrano, B.R. Heckel, and E.G. Adelberger.
\newblock {\em Class. Quantum Grav.}, 28:145011, 2011.

\bibitem{Landau}
S.J. Landau, F.A. Teppa-Pannia, Y.~Bonder, and D.~Sudarsky.
\newblock {\em Astroparticle Physics}, 35:377, 2012.

\bibitem{matter}
T.K. Kuo and J.~Pantaleone.
\newblock {\em Rev. Mod. Phys.}, 61:937, 1989.

\bibitem{MSW1}
L.~Wolfenstein.
\newblock {\em Phys. Rev.}, D17:2369, 1978.

\bibitem{MSW2}
S.P. Mikheyev and A.Y. Smirnov.
\newblock {\em Nuovo Cimento}, 9C:17, 1986.

\end{thebibliography}

\end{document}